\documentclass[prd,superscriptaddress,floatfix,amsmath,footinbib,amssymb,twocolumn]{revtex4}
\usepackage{amssymb}
\usepackage{amsmath}
\usepackage{amsfonts}
\usepackage{tikz}
\usepackage{bm}

\usepackage{epsfig}
\usepackage{t1enc}
\usepackage{soul}
\usepackage{color}

\usepackage{mathrsfs}
\usepackage{url}
\usepackage[all]{xy}
\usetikzlibrary{calc,patterns,angles,quotes}

\begin{document}

\title{Reply to the comment on "Quantum principle of relativity"}

\begin{abstract}

We address concerns raised by Horodecki \cite{Horodecki2023} towards our work on the relation between the extended relativity and fundamental aspects of quantum theory.

\end{abstract}

\author{Andrzej Dragan}
\affiliation{Institute of Theoretical Physics, University of Warsaw, Pasteura 5, 02-093 Warsaw, Poland}
\affiliation{Centre for Quantum Technologies, National University of Singapore, 3 Science Drive 2, 117543 Singapore, Singapore}

\author{Artur Ekert}
\affiliation{Centre for Quantum Technologies, National University of Singapore, 3 Science Drive 2, 117543 Singapore, Singapore}
\affiliation{Mathematical Institute, University of Oxford, Woodstock Road, Oxford OX2 6GG, United Kingdom}
\affiliation{Okinawa Institute of Science and Technology, Onna, Okinawa 904-0495, Japan}

\begin{abstract}
We discuss critical remarks raised by Horodecki \cite{Horodecki2023} towards our work on the connection between superluminal extension of special relativity and fundamental aspects of quantum theory.
\end{abstract}

\maketitle

In a recent paper \cite{Dragan2020}, we demonstrated that extending special relativity to superluminal frames of reference inevitably leads to fundamental indeterminacy known from quantum theory. Furthermore, we argued that within this extension, motion along multiple paths also becomes inevitable, drawing parallels with quantum superpositions. Finally, we established that a probabilistic and covariant description of such motions necessitates the use of complex probability amplitudes.

These claims elicited several concerns by Grudka and Wójcik \cite{Grudka2022} as well as Del Santo and Horvat \cite{DelSanto2022}, which we have previously addressed \cite{Dragan2022a, Dragan2022b}. Recent questions and critiques posed by Horodecki \cite{Horodecki2023} are addressed below. 

Horodecki's main concern questions whether extended special relativity alone can be used to deduce quantum theory in its entirety, or if the proposed "quantum principle of relativity" remains incomplete in this regard. Our short answer is that we don't know yet. While we've demonstrated that some of the most profound properties of quantum theory, such as its inherent randomness and the superposition principle, can be deduced from extended relativity, this doesn't imply that all formal aspects of the theory can be similarly derived. In fact, we've illustrated that the probability amplitudes formalism represents merely the simplest covariant solution, leaving room for other potential descriptions.

Horodecki references Heisenberg's classification of possible universes based on the values of physical constants $\frac{1}{c}$ and $\hslash$. Specifically, he considers the scenario where $\frac{1}{c} = 0$ and $\hslash \neq 0$, representing a quantum but not relativistic model of reality, arguing that the universe could be quantum, without being relativistic. However, taking a low-energy limit of Dirac's theory and arriving at the approximate Pauli equation does not mean that the resulting theory is truly non-relativistic. In our work we have argued that the reason we have to consider probabilistic description involving superpositions is due to relativity. At this stage it is secondary, whether the dynamical equation is strict, or only approximate. It is best illustrated by the fact that the "non-relativistic" Pauli theory still involves spin with the gyromagnetic factor $g=2$ which is truly relativistic. It is also in principle possible to imagine a universe, in which the speed of light is infinite. However this does not invalidate our claims, that quantum effects are a consequence of relativity, either. To show that, let us draw an analogy. Elliptical trajectories of planets arise from Newton's law of gravitation. However, it's conceivable for a universe to exhibit such trajectories even in the absence of gravity, driven by entirely different, alternative laws of physics. Obviously, this doesn't detract from the earlier assertion that Kepler's laws directly originate from Newtonian gravity. It is conceivable that in an alternative universe quantum theory might have emerged "out of the blue". However, our reasoning suggests that, in our universe, several fundamental aspects of quantum theory are a consequence of relativity. This can also lead to a speculation that fundamental constants may not be entirely independent, in particular $\lim_{c\to\infty}\hslash =0$.

Another observation made by Horodecki is that there are fundamentally two types of indeterministic events in quantum theory: those due to spontaneous decay processes, and those that occur within the process of measurement. Horodecki writes, "We have no convincing evidence that measurement randomness can always be attributed to the randomness associated with particle decays." If we interpret "particle decays" more broadly as particle interactions (considering that for tachyons a decay can be Lorentz-transformed into a collision), this statement becomes contentious. Presently, our understanding of physics hinges on elementary interactions within the standard model. Beyond gravity, there are no other known forces that could account for the emerging laws of physics. For example, photo-avalanche detectors produce unpredictable readings, initiated with an interaction between a single photon and a single electron, leading to a microscopic electric current which is then amplified by the detection mechanism. Consequently, there is no reason to reject the claim that, according to the present understanding of physics, quantum unpredictability can be traced back to the unpredictability of individual particle interactions. And, as we argued, the unpredictability of the latter can be understood using relativistic arguments.

The next question raised by Horodecki is whether the laws of physics, such as the Born rule, can remain consistent in all reference frames, including superluminal ones, in the 1+3 dimensional case, when the former and the latter can be distinguished (not being fully invariant). The ability to differentiate between conventional (subluminal) reference frames is not a concern when considering non-inertial frames of reference. For instance, in the Rindler frame, which corresponds to a relativistically uniformly accelerated observer, the Born rule continues to apply seamlessly. Distinguishability of that frame from inertial frames is not a problem. Hence, the fact that superluminal inertial observers employ a different metric than their subluminal counterparts should not present any problem neither. In fact, general covariance, which is the foundation of general relativity, states that the laws of physics should not be affected by the choice of coordinates. The quantum principle of relativity that we proposed simply extends this domain of possible coordinate systems to include superluminal observers.

Horodecki notices that in the derivation of the expressions for complex probability amplitudes we only employ conventional, subluminal covariance between frames. This is true. However the questions that have to be answered first are: why do we need to look for a probabilistic description of particle dynamics, and why do we need to consider motion along multiple paths in the first place? These questions are justified only by considering superluminal frames of reference. And they provide us with a justification to look for relativistically invariant probability distributions. 

Finally, Horodecki raises a crucial and well-founded critique: does our proposal yield any measurable and potentially observable effects? The straightforward answer is: if tachyons existed, then this would undoubtedly be the case. It is worth noting that we have recently demonstrated that a covariant quantum field theory of tachyons with a positive-definite spectrum and a stable, invariant vacuum can be constructed \cite{Paczos2023}. In this study, as well as in our previous work \cite{Dragan2023}, we emphasize that the Higgs mechanism incorporates tachyonic fields. As a result, our ongoing research opens the way to study fully quantized theory of spontaneous symmetry breaking. This leaves us with a hope that the answer to the last question posed by Horodecki is affirmative.


\begin{thebibliography}{9}

\bibitem{Dragan2020}
A. Dragan and A. Ekert, New J. Phys. {\bf 22}, 033038 (2020).

\bibitem{Grudka2022}
A. Grudka and A. Wójcik, New J. Phys. {\bf 24}, 098001 (2022).

\bibitem{DelSanto2022}
F. Del Santo and S. Horvat, New J. Phys. {\bf 24}, 128001 (2022).

\bibitem{Dragan2022a}
A. Dragan and A. Ekert, New J. Phys. {\bf 24}, 098002 (2022).

\bibitem{Dragan2022b}
A. Dragan and A. Ekert, New J. Phys. {\bf 25}, 088002 (2022).

\bibitem{Dragan2023}
A. Dragan, K. Dębski, S. Charzyński, K. Turzyński, and A. Ekert, Class. Quantum Grav. {\bf 40}, 025013 (2023).

\bibitem{Paczos2023}
J. Paczos, K. Dębski, S. Charzyński, K. Turzyński, A. Ekert, and A. Dragan, arXiv:2308.00450 [quant-ph] (2023).

\bibitem{Horodecki2023}
R. Horodecki, arXiv 2301.07802 [quant-ph] (2023).

\end{thebibliography}
\end{document}